\documentstyle[12pt,epsf]{article}
\newcommand{\CC}{{\cal C}}
\newcommand{\CP}{{\cal P}}
\newcommand{\tCC}{{\tilde{\CC}}}
\newcommand{\tCP}{{\tilde{\CP}}}
\newcommand{\CR}{\nonumber \\}

\renewcommand{\thefootnote}{\fnsymbol{footnote}}
\def\pslash{\mathord{\not\mathrel{p}}}
\def\tilpslash{\mathord{\not\mathrel{\tilde p}}}
\def\Pislash{\mathord{\not\mathrel{\Pi}}}
\def\Pslash{\mathord{\not\mathrel{P}}}

\def\tbarmu{\bar{\theta} \Gamma_{\mu} \dot{\theta}}
\def\tbarnu{\bar{\theta} \Gamma_{\nu} \dot{\theta}}
\def\half{\frac{1}{2}}
\def\thetabar{\bar{\theta}}
\def\del{\partial}
\def\lb{\newline}

\makeatletter
\@addtoreset{equation}{section}

\makeatother

\begin{document}

\begin{titlepage}
\null
\begin{flushright}
FERMILAB-Pub-00/155-T 
\\
TIT/HEP-434
\end{flushright}

\vskip 2cm
\begin{center}
{\LARGE \bf D0 and D1 Branes With $\kappa_-$ and $\kappa_+$ Extended}
\vskip .5cm
{\LARGE \bf Symmetry}
\lineskip .75em
\vskip 2.5cm

\normalsize

  {\large \bf Moshe Moshe~$^{2,3}$}
\footnote{\it  e-mail address: 
moshe@physics.technion.ac.il}  ~and~~ {\large \bf 
Norisuke Sakai~$^{1}$}
\footnote{\it  e-mail address: nsakai@th.phys.titech.ac.jp}

\vskip 1.5em

{ \it   $^{1}$Department of Physics, Tokyo Institute of Technology\\
Tokyo 152-8581, JAPAN \\ $^{2}$Department of Physics , Technion -
 Israel Institute of Technology 
\\ Haifa 32000, ISRAEL \\ $^{3}$Fermi National Accelerator Laboratory\\
P.O. Box 500, Batavia, Illinois 60510, USA }

\vskip 2cm

{\bf Abstract}

\end{center}

D0 brane (D-particle) and D1 brane actions 
possess first and second class constraints that result in 
local $\kappa$ symmetry. The $\kappa$ symmetry of the 
D-particle and the D1 brane is extended here into a larger symmetry
($\kappa_-$ and $\kappa_+$) in a larger phase space 
by turning second class constraints into first class. 
Different gauge fixings of these symmetries result in 
different presentations of these systems while a "unitary" 
gauge fixing of the new $ \kappa_+$ symmetry retrieves the
original  action with $\kappa_- = \kappa $ symmetry. For D1 brane our 
extended phase space makes all constraints into first 
class in the case of vanishing world sheet electric field (namely $(0, 1)$ 
string). 

\end{titlepage}

\renewcommand{\thefootnote}{\arabic{footnote}}
\baselineskip=0.7cm

\clearpage

{ \bf Introduction}

\vspace{1cm}

An important ingredient in the study of D-branes\cite{POLCH}
dynamics is their local
fermionic symmetry on the world-volume, the $\kappa$ symmetry.
The history of this symmetry goes back to 
the superparticle action\cite{BRINK}
where it was identified \cite{AZCA}\cite{SIEGE1} and applied
to the superstring\cite{GREEN}. It was used also in the study
of super p-branes\cite{HUGHE} in different dimensions.
The role of 
the $\kappa$ symmetry 
was further emphasized in the
 study of the D-branes embedded in flat 10D space-time
in \cite{AGANA}\cite{BERGS}. The symmetry is generated by 
16 irreducible first class fermionic constraints. 
These constraints are accompanied by another
set of 16 second class fermionic constraints 
which do not correspond to
any local symmetry. The covariant separation of the two
types of constraints in the brane action 
was emphasized in \cite{KALLO}\cite{KAMIM} and  enabled the
covariant quantization of the D0 and D1 brane. 

It has been found difficult to quantize covariantly
 the massless superparticle, as is the situation also with
 the Green-Schwarz formulation of the superstring 
\cite{GREEN} since in both systems first and second class constraints
cannot be separated in a covariant manner.
 This is a long lasting problem and many attempts have been made
to solve it \cite{HORI}-\cite{SIEGE2}.
In the massive superparticle action the 
$\kappa$ symmetry is explicitly broken. Its first class
constraints are replaced now by solvable second class 
constraints and the system can be quantized 
covariantly by means of Dirac brackets 
since all its constraints are
second class.  
Since the massive superparticle can be quantized covariantly, one may be 
tempted to consider the massless limit of the massive case as a substitute 
for the covariant quantization of the massless superparticle.
However, the Dirac brackets become singular in the  $p^2=m^2 \to 0$ limit. 
The restoration of the broken
$\kappa$ symmetry of the massive system in an 
extended phase space \cite{FADDE}\cite{BATAL} by adding extra fermionic degrees of 
freedom was considered in\cite{FEINB}. 
Another possibility to restore the $\kappa$ symmetry is to include a 
 proper Wess-Zumino term in the action, as is the case with the D0 brane
 \cite{AGANA}-\cite{KAMIM}. 
This is physically more interesting, but contains
in addition to the first class constraints, 
that correspond to the restored $\kappa$ symmetry,
also second class constraints . 
When considering the massless limit, one finds the need to avoid 
these second class constraints since also here the Dirac brackets 
become singular in the massless limit.
The restoration of symmetry with no second class
constraints, gives the full advantages
of working within a system with local symmetry in particular 
 a covariant wave function can be formulated
also in the 
massless limit\cite{FEINB}.  For this purpose, 
it is usually useful to turn the second 
class constraints into first class. 
This formulation offers a flexibility to allow various gauge fixings which
are physically equivalent. 
At the same time, the newly introduced first class constraints generate 
a gauge symmetry which may give more insight into the geometrical 
structure of the system which is interesting in its own right.

Several other different approaches to this issue share in 
common the idea of adding extra 
dynamical degrees of freedom while extending the symmetry of the system 
in different manners. 
In the 
geometrical-superembedding approach,
superbrane dynamics
is manifestly supersymmetric on the worldvolume as well as in  target
superspace\cite{SOROK} and the auxiliary 
commuting spinors superpartners have 
twistor-like and Lorentz harmonics properties.
This approach, which has a wide range of applications
in several physical systems,
has been developed for super p-branes and
D branes as well. Other treatments of second class constraints include
extended phase space variables in \cite{EISEN} and, more recently, auxiliary
commuting twistor-like spinor variables and tensorial central charge coordinates 
were used in \cite{BANDO}.
Introducing Liouville 
mode while solving 
the second class constraints left a final action with
only first class constraints in \cite{SIEGEb}. Other related approaches
can be found in\cite{BERKO}-\cite{GALAJ}.

In the first part of this paper we
suggest a new symmetric system for the D-particle 
in which
the second class constraints are turned into first class in
an extended phase space which includes extra fermionic 
degrees of freedom.
We define a system that
contains
$\theta_\alpha,\pi_\alpha$, the original fermionic degrees of freedom
of the D0 brane to which extra fermionic degrees of freedom
$\zeta_\alpha,\rho_\alpha$ are added  (  $\zeta_\alpha ,
\rho_\alpha$ are
Majorana-Weyl spinors while $\theta_\alpha,
\pi_\alpha$  are only Majorana).
The new system has, in addition to the original $\kappa = \kappa_-$
symmetry a new local $\kappa_+$ symmetry. The system can be 
gauge fixed in many different ways while one of these gauge
fixings ("unitary" gauge) retrieves the original D0 brane.
The rest of the paper presents, along the same lines,
 the D1 brane with an extended $\kappa_-$
and $\kappa_+$ local symmetry. We consider the case of a
vanishing electric field in the Born-Infeld-Nambu-Goto action. 

\eject

{ \bf  Superparticle and D-particle}

\vspace{.5cm}

The N=1 massless superparticle action in d=10 space-time dimensions
(\cite{BRINK}-\cite{GREEN}):
\begin{eqnarray}
S = \int^{\tau_f}_{\tau_i}   {\cal L}(\tau) d\tau 
= - \frac{1}{2}\int^{\tau_f}_{\tau_i}  d\tau {1\over e}
( \dot{x}^{\mu} - i \thetabar_+\Gamma^{\mu} \dot{\theta_+} )^2
\label{eq:SuperP-action}\end{eqnarray}
is invariant under the local $\kappa$ symmetry:
\begin{eqnarray}
\delta x_\mu = i \thetabar_+\Gamma_{\mu} \delta{\theta_+} \ \ , \ \ 
\delta\theta_+ =( \dot{x}^{\nu} - i \thetabar_+
\Gamma^\nu\dot{\theta}_+ )\Gamma_\nu\kappa_- \ \ ,\ \
\delta e = 4 i e \dot{\thetabar}_+\kappa_-
\label{eq:kappa-transf}\end{eqnarray}
$x^{\mu}~~ (\mu = 0, 1...9) $ 
are space-time coordinates and
$ \theta_{+}$ is a Majorana Weyl spinor with positive 
(or negative) chirality, the spinor $\kappa_-$ has
the opposite chirality of $ \theta_{+}$
and $e(\tau)$ is the "einbein" of 
local reparametrization symmetry. 
The $32\times32~$ $\Gamma^\mu$ matrices  $(\mu=0,1,2 .. ,9)$ 
are built out of the conventional spin(8) 
matrices\footnote{
Our conventions are:
$\Gamma^{m} =\sigma^1\otimes \gamma^m 
\ , \ m=1,2, . . 9
\ , \ \Gamma^{0} =-i\sigma^2 \otimes {\cal I} $
$\ , \  
\Gamma^{11} =\sigma^3 \otimes{\cal I}  
 \ , \  \{\Gamma^{11},\Gamma^\mu\}= 0
\ , \ $
  $\Gamma^\mu =
\left( \matrix{ 0 & \bar\gamma^\mu \cr \gamma^\mu & 0 } \right) 
\ \ , \ \  \mu = 0,1,2, . . .9 \ \ \  
\bar\gamma^\mu = \{ -1 , \gamma^l \} \ \ , \ \ \gamma^\mu = \{ 1 , 
\gamma^l \}
\ \ ,\ \ l=1,2, . . 9  $
$
\gamma^k \ \  \{k=1,2, . . 8\}  
\mbox{ are $16\times 16$ spin(8) matrices } \ , \ 
 \gamma^9=\Pi_{k=1}^{k=8}\gamma^k \ , \ $  
$\{ \gamma^m,\gamma^n \} = 2\delta^{m,n}  \ , \ m,n=1,2..9 $
$\bar\gamma^\mu \gamma^\nu + \bar\gamma^\nu \gamma^\mu  =
\gamma^\mu \bar\gamma^\nu + \gamma^\nu  \bar\gamma^\mu 
=2\eta^{\mu\nu}
 \ , \ \mu,\nu=0,1,2..9 $
} and satisfy 
$\{\Gamma^\mu,\Gamma^\nu\}=2\eta^{\mu\nu}$ and 
$\eta^{\mu\nu}=diag\{-+++..\}$

The system has 8 fermionic 
first class constraints
and 8 fermionic second class constraints and thus its phase space has 
$~(32-2\times8-8)~$ 8 independent fermionic
degrees of freedom. 

Local $\kappa$ symmetry is explicitly broken in the 
N=1 massive superparticle action in d=10 dimensions \cite{FEINB}:
\begin{eqnarray}
S = \int^{\tau_f}_{\tau_i}   {\cal L}(\tau) d\tau 
= \int^{\tau_f}_{\tau_i}  d\tau \{ - {1\over {2e}}
( \dot{x}^{\mu} - i \bar{\theta_+} \Gamma^{\mu} \dot{\theta_+})^2 
+ \half e m^2 \}
\label{eq:massive SP}\end{eqnarray}
Here, using Eq.(\ref{eq:kappa-transf}), one finds 
$\delta{\cal L} = 2 i 
e m^2 \dot{\thetabar}_+\kappa_- \neq 0$. All 16
constraints are second class and
its phase space has $(32-16=) 16$ independent fermionic degrees of freedom.

One possible modification by which the
local $\kappa_-$ symmetry can be restored is extending 
its phase space  to N=2 
while adding an appropriate Wess-Zumino term. 
\begin{eqnarray}
\delta x_\mu = i\thetabar \Gamma_{\mu} \delta{\theta} \ \ , \ \ 
\delta\theta_+ =( \dot{x}^{\nu} - i\thetabar
\Gamma^\nu\dot{\theta} )\Gamma_\nu\kappa_- \ \ ,\ \
\delta e = 4i e \dot{\thetabar}_+\kappa_-  
\label{eq:kappa-transf-2}\end{eqnarray}
\begin{eqnarray}
{\cal L} =
 -{1\over {2e}}
( \dot{x}^{\mu} - i \bar{\theta} \Gamma^{\mu} \dot{\theta})^2 
+ \half e m^2 + {\cal L}_2
\label{eq:Lagrangian-massive}\end{eqnarray}
Here $\theta = \theta_+ + \theta_- $ ($\theta$ is a Majorana spinor
and $\theta_+$ and $\theta_-$ are Majorana-Weyl spinors of
opposite chirality) 
and $\delta \theta_-$ and ${\cal L}_2$ are to be determined below. 
{}From Eq.(\ref{eq:kappa-transf-2}) one finds:
\begin{eqnarray}
\delta{\cal L} = \frac{2i}{e}
( \dot{x}^{\mu} - i \bar{\theta} \Gamma^{\mu} \dot{\theta})^2 
\dot{\thetabar}_+\kappa_- - \frac{2i}{e}
( \dot{x}^{\mu} - i \bar{\theta} \Gamma_{\mu} \dot{\theta})
\dot{\thetabar}\Gamma^\mu (\delta\theta_+ + \delta\theta_-)
+ 2 i e m^2\dot{\thetabar}_+\kappa_- + \delta{\cal L}_2
\label{eq:Lagrangian-transf}\end{eqnarray}
$\delta{\cal L}=0$ for a properly chosen ${\cal L}_2$. 
A possible solution of the form:
\begin{eqnarray}
\delta{\cal L}_2 = A_+\delta\theta_- + B_-\delta\theta_+ 
\label{eq:Lagrangian2-transf}\end{eqnarray}
gives $A_+=-2 i m\dot{\thetabar}_+ \ , \ B_-=2 i m \dot{\thetabar}_-$
and  $\delta\theta_-=em\kappa_-$
(up to a rescaling $A_+\to A_+/\alpha \ , \ B_-\to B_-\alpha$
and $\delta\theta_-\to \delta\theta_-\alpha$)
\begin{eqnarray}
\delta{\cal L}_2 = i m\delta(\dot{\thetabar}\Gamma^{11}\theta)
 - i m{d\over {dt}}(\delta\thetabar\Gamma^{11}\theta) 
\label{eq:Lagrangian-2-solution}\end{eqnarray}
where
\begin{eqnarray}  
\Gamma^{11} =\sigma^3 *{\cal I}  = \left( \matrix{ 1 & 0 \cr
0 & -1 } \right) \ \ \ , \ \ \{\Gamma^{11},\Gamma^\mu\}= 0
\nonumber
\end{eqnarray}
(${\cal I}$ is the $16\times 16$  identity matrix)
Thus,
\begin{eqnarray} 
{\cal L}(\tau) = -\frac{1}{2}e^{-1}
( \dot{x}^{\mu} - i \bar{\theta} \Gamma^{\mu} \dot{\theta} )^2 + 
\half e m^2  - i m\bar{\theta} \Gamma^{11} \dot{\theta}  
\label{eq:Lagrangian WZ}\end{eqnarray}
has a restored $\kappa_-$ symmetry. The system has now
not only 16 first class constraints but also  16 second class constraints
and the number of independent degrees of freedom in phase space 
is the same as the N=1 massive superparticle
($64-2\times 16-16=16$). Indeed, when compared to the massive N=1
superparticle action in Eq.(\ref{eq:massive SP}), 
the added negative chirality $\theta_-$ degrees of
freedom ( 32 degrees of freedom in phase space;
$\theta_-$ and their canonical conjugate $\bar\pi_+$)
can be gauged away once the restored $\kappa_-$ symmetry
is gauge fixed $( \theta_- = 0)$. 
One is left, after gauge fixing, back with ${\cal L}(\tau)$
of the massive N=1 superparticle in Eq.(\ref{eq:massive SP}).

A very appealing point of view on ${\cal L}(\tau)$ of
Eq.(\ref{eq:Lagrangian WZ})   
is obtained when one starts with the massless superparticle 
action in d=11 dimensions which is given by 
(\cite{BRINK}-\cite{GREEN}):
\begin{eqnarray}
S = \int^{\tau_f}_{\tau_i}   {\cal L}(\tau) d\tau 
= - \frac{1}{2}\int^{\tau_f}_{\tau_i}  d\tau e^{-1}
(\dot{x}^{\hat m} - i \bar{\theta} \Gamma^{\hat m} \dot{\theta})^2
\label{eq:Action-d11}\end{eqnarray}
where $x^{\hat m}~~ ({\hat m} = 0, 1...10) $ are the space-time coordinates
and
$ \theta_{\alpha}=\theta_{+\alpha}+\theta_{-\alpha} (\alpha=1,2.  ..32)$ 
are the corresponding
fermionic coordinates which can be regarded as two Majorana Weyl 
spinors of opposite chiralities, if viewed as spinors in ten dimensions. 

When one of the space directions is compactified
 \cite{GREEN2} to 
a radius of $R=m^{-1}=Z^{-1}$, the d=11 massless
superparticle action results \cite{KALLO}\cite{KAMIM} 
in the D0 brane action. 
\begin{eqnarray}
S = \int^{\tau_f}_{\tau_i} {\cal L}(\tau) d\tau= \int^{\tau_f}_{\tau_i} 
d\tau \{\!\!\!\!\!& - &\!\!\!{1\over{2e}}
( \dot{x}^{\mu} - i \bar{\theta} \Gamma^{\mu} \dot{\theta} )^2 + 
\half e Z^2 - i Z \bar{\theta} \Gamma^{11} \dot{\theta} ~\}  \nonumber \\
&+&\!\!\!Z[x_{10}(\tau_f)-x_{10}(\tau_i)]
\label{eq:Action-compactified}\end{eqnarray}
Where $p_{10}$ was set to $p_{10}=m=Z$ , $\Gamma^{\hat{10}} $ is defined as 
 $\Gamma^{11} $ and ${\mu} = 0, 1...9 $.

The D0 brane action in Eq.(\ref{eq:Action-compactified}) is the same action 
obtained in Eq.(\ref{eq:Lagrangian WZ}) and its
Wess-Zumino term $Z \bar{\theta} \Gamma^{11} \dot{\theta}$ 
establishes the local $\kappa_-$ symmetry, which is the
original symmetry of the d=11 massless superparticle 
action.
Thus, instead of 32 second class constraints as in the N=2, d=10 massive
superparticle action,  
the D0 has 16 first class constraints and 16 second class constraints 
which is the same number of constraints as 
the massless  N=2, d=10 superparticle and here too the 16 first
class constraints result in $\kappa_-$ symmetry. 
An important difference between the
D0 action and the massless superparticle is the fact that in the D0 case the
first and second class constraints can be separated in a covariant 
manner\cite{KALLO}\cite{KAMIM}, this cannot be done for the massless N=2
d=10 superparticle.

We would like to treat now the D0 system in a more symmetrical
manner by turning also its remaining 16 second class constraints into
first class. The resulting system will have in addition to the
original $\kappa_-$ symmetry also a $\kappa_+$ symmetry
generated by the new first class constraints.
Among all possible different  gauge fixing of the new
$\kappa_+$ symmetry, one should also 
be able to retrieve the original D0 system, by appropriately
gauge fixing ("unitary" gauge fixing) the extended symmetric
system.

After implementing the $\kappa_+$  extended symmetry into the system,
the number of independent degrees of freedom should not change.
Thus, one has to extend the phase space of the new, symmetric system 
by adding extra 
fermionic degrees of freedom to account for the
increase of symmetry. In the following we define and summarize the
properties of the $\kappa_+ , \kappa_-$ symmetric system.
{}From Eq.(\ref{eq:Action-compactified})(ignoring the boundary term) or from: 
\begin{eqnarray}
S = -Z\int^{\tau_f}_{\tau_i} d\tau \{
\sqrt { -( \dot{x}^{\mu} - i \bar{\theta} \Gamma^{\mu} \dot{\theta})^2  } + 
 i \bar{\theta} \Gamma^{11} \dot{\theta} ~\} 
\label{eq:Action-sqrt-WZ}\end{eqnarray}
one finds the constraints,
\begin{eqnarray}
\bar T_\alpha = \bar \pi_\alpha + i (\thetabar\pslash)_\alpha + 
i Z(\thetabar\Gamma^{11})_\alpha = 0 \ \ , \ \ p^2 + Z^2 = 0 
\label{eq:constraints-1}\end{eqnarray}
where $\bar\pi_\alpha$ is the momentum, canonical conjugate 
of $ \theta_\alpha $ 
(right handed derivatives are used when taking a derivative
with respect to $\dot\theta_\alpha$). 

The momentum is: 
$p_\mu=Z \{ { {\dot x_\mu - i \tbarmu} \over {\sqrt{ 
 -(\dot x_\nu - i \tbarnu)^2}}}\} $ ; 
the Hamiltonian is $ H_0=0$. Using the Poisson brackets
\begin{eqnarray}
[x_{\mu},p^{\nu}] = \delta^{\nu}_{\mu}, \ \ \ \
[\theta^{\alpha},
\bar\pi_{\beta}] = \delta^{\alpha}_{\beta} 
\ \ \ \mbox{all others} \ = 0 ~~.
\label{eq:Commutation-Relations}\end{eqnarray}
one finds:
\begin{eqnarray}
[\bar T_\alpha , \bar T_\beta ] 
= 2 i (\Gamma^0 (\pslash + Z\Gamma^{11}))_{\alpha\beta}
\label{eq:Constraints-PB}\end{eqnarray}
and we have: 
\begin{eqnarray}
\pslash + Z\Gamma^{11}= \left(  \matrix{ Z & \bar{\pslash} \cr
\pslash & -Z }  \right) 
 \ \ \mbox{and } \ \ (\pslash + Z\Gamma^{11})^2=(p^2+Z^2)*{\cal I}
\label{eq:pslash+Z}\end{eqnarray}
(here, ${\cal I}$ is the $32\times 32$ identity matrix)
\begin{eqnarray}
{\rm det}[\Gamma^0(\pslash + Z\Gamma^{11})]=(p^2+Z^2)^{16}=0 .
\label{eq:Det}\end{eqnarray}
In the $32\times 32$ matrix 
$\Gamma^0(\pslash + Z\Gamma^{11})$, 
each of its
$16\times 16$ blocks has a non-zero determinant, and 
$\Gamma^0(\pslash + Z\Gamma^{11})$ has rank
16 . The first and second class
constraints can be covariantly separated by defining
\cite{KALLO}\cite{KAMIM}:
\begin{eqnarray}
&\bar T_1 = \bar T  (\pslash + Z\Gamma^{11})( { {1-\Gamma^{11}}\over {2}} )
= \bar\pi_-\pslash - Z\bar\pi_+ + i \thetabar_+(p^2+Z^2) \\ \nonumber
\ \ \ \mbox{and}& \ \ \ \bar T_2 = \bar T ({ {1+\Gamma^{11}}\over {2}})
= \bar\pi_-  + i \thetabar_+ \pslash 
 + i Z \thetabar_{-}
\label{eq:T1 and T2}\end{eqnarray}
as seen from the following Poisson bracket relations:
\begin{eqnarray}
[\bar T_{1\alpha} , \bar T_{1\beta} ] 
= -2 i (p^2+Z^2)(\Gamma^0 { {1+\Gamma^{11}}\over {2}}
\pslash)_{\alpha\beta} 
\ \ \ , \ \ \ [\bar{T}_{2\alpha} , \bar{T}_{2\beta} ] 
= 2 i (\Gamma^0 { {1-\Gamma^{11}}\over {2}}
\pslash)_{\alpha\beta} \nonumber
\end{eqnarray}
\begin{eqnarray}
[\bar T_{1\alpha} , \bar T_{2\beta} ] 
= -2 i (p^2+Z^2)(\Gamma^0 { {1+\Gamma^{11}}\over {2}}
)_{\alpha\beta}
\label{eq:Constraints-PB2}\end{eqnarray}
where we used:
\begin{eqnarray}
[\theta_{\pm\alpha},\bar\pi_{\mp\beta}] =
({ {1\pm\Gamma^{11}}\over {2}} )_{\alpha\beta} .
\label{eq:PB+-}\end{eqnarray}
The generator of $\kappa$ symmetry and reparametrization 
is given in terms of the parameters $\kappa_-$ and $\epsilon_p$ by:
\begin{eqnarray}
G={1 \over 2}\epsilon_p(p^2+Z^2) + \bar T_1\kappa_-
\label{eq:Generator-1}\end{eqnarray}
As mentioned above, the D0 brane has a total of 16    
independent fermionic degrees of
freedom in phase space $(~32\times2-(2\times16+16)~)$ 
as reflected by the 16 first class and 16 second class
fermionic constraints in Eq.(\ref{eq:T1 and T2}).

In an extended phase space where the system is described
by extra degrees of freedom, second class constraints can be turned into
first class\cite{FADDE}\cite{BATAL}.
One denotes the second class constraints Poisson bracket by:
\begin{eqnarray}
[\bar{T}_{2\alpha} , \bar{T}_{2\beta} ] 
= 2 i ( \Gamma^0 { {1-\Gamma^{11}}\over {2}}
\pslash )_{\alpha\beta} = V_{\alpha\gamma}
\omega_{\gamma\delta}V_{\delta\beta}
\label{eq:T2-PB}\end{eqnarray}
 $V_{\alpha\beta}$ constructs the BRST operator in the extended symmetric
system and $\omega_{\gamma\delta}$ is used in order to 
define a linear 
combination of extra
32 fermionic degrees of freedom in phase space. We have (up to similarity
transformations of $\omega$ in the symplectic structure of 
Eq.(\ref{eq:T2-PB})):
\begin{eqnarray}
V_{\alpha\beta}=\left( \Gamma^0 { {1-\Gamma^{11}}\over {2}}
\pslash\right)_{\alpha\beta} \ \ , \ \ \omega_{\alpha\beta} 
= -{2 i \over {p^2}}\left(
({ {1+\Gamma^{11}}\over {2}} ) \pslash\Gamma^0\right)_{\alpha\beta}
\label{eq:V and Omega}\end{eqnarray}
We define the linear combination:
\begin{eqnarray}
\bar\Phi_{-\alpha}=-\half\bar\rho_{-\alpha}
+ \tilde\omega_{\alpha\beta}\zeta_{+\beta} =-\half\bar\rho_{-\alpha}
 - i \half(\bar\zeta_{+}\pslash)_{\alpha}
\label{eq:Linear Comb.}\end{eqnarray}
where we used
\begin{eqnarray}
\tilde\omega_{\alpha\beta}= -i \half \left(
\Gamma^0({ {1-\Gamma^{11}}\over {2}} ) \pslash\right)_{\alpha\beta}, \ \ \ \ 
\
\omega_{\alpha\gamma}\tilde\omega_{\gamma\beta}=
\left({ {1+\Gamma^{11}}\over {2}}\right)_{\alpha\beta}
\label{eq:Omega}\end{eqnarray}
$\rho_-$ and $\zeta_+$
are a canonical pair of Majorana-Weyl spinors representing 
extra 32 fermionic degrees of freedom
whose Poisson bracket is:
\begin{eqnarray}
[\bar\rho_{-\alpha},\zeta_{+\beta}] =
\left( { {1+\Gamma^{11}}\over {2}}\right)_{\alpha\beta}
\label{eq:rho zeta PB}\end{eqnarray}
The linear combination in Eq.(\ref{eq:Linear Comb.})
 of the extra degrees of freedom $\Phi_{-\alpha}$
have the Poisson bracket:
\begin{eqnarray}
[\bar\Phi_{-\alpha},\bar\Phi_{-\beta}] =-\tilde\omega_{\alpha\beta}=
 i \half\left(\Gamma^0
({ {1-\Gamma^{11}}\over {2}} ) \pslash\right)_{\alpha\beta}
\label{eq:PHI-PHI-PB}\end{eqnarray}
One defines now, in the extended phase space, the
following constraints, which are first class:
\begin{eqnarray}
\bar T '_{-\alpha}= \bar T_{2\alpha} + 
\bar\Phi_{-\beta}\omega_{\beta\gamma}V_{\gamma\alpha} \ \ \ \ , \ \ \ \
[\bar T'_{-\alpha},\bar T'_{-\beta}] = 0
\label{eq:New Constraints}\end{eqnarray}

Thus, the dynamics in the extended phase space are defined by the 
two opposite chirality sets of constraints $\bar T_{+}, \bar T'_{-}$
 and their Poisson bracket:
\begin{eqnarray}
\bar T_{1}\equiv\bar T_{+}=\bar\pi_-\pslash  
 - Z\bar\pi_+ + i \thetabar_+(p^2+Z^2)\ \ , \ \
\bar T'_{-}=\bar\pi_-  + i \thetabar_+ \pslash 
 + i Z \thetabar_{-} 
 - i \bar\rho_- + \bar\zeta_+\pslash \ \  .
\label{eq:T+-}\end{eqnarray}
Using the extended phase space Poisson brackets in
Eq.(\ref{eq:PB+-}) and Eq.(\ref{eq:rho zeta PB})
one finds the Poisson brackets of two chiral multiplets of first class
constraints :
\begin{eqnarray}
[\bar T_{+\alpha},\bar T_{+\beta}] = -2 i (p^2+Z^2)\left(\Gamma^0
({ {1+\Gamma^{11}}\over {2}} ) \pslash\right)_{\alpha\beta} \ \ , \ \
[\bar T'_{-\alpha},\bar T'_{-\beta}] = 0   \nonumber
\end{eqnarray}
\begin{eqnarray}
[\bar T_{+\alpha},\bar T'_{-\beta}] = -2 i (p^2+Z^2)\left(\Gamma^0
({ {1+\Gamma^{11}}\over {2}} )\right)_{\alpha\beta}
\label{eq:T+-NEW PB}\end{eqnarray}
The total extended phase space hamiltonian is:
\begin{eqnarray}
H_T = H_0 + \half\lambda_p(p^2+Z^2)
 + \bar T_+ \lambda_- + \bar T'_-\lambda_+  
\ \ \ , \ \ \ H_0 = -\half e(p^2+Z^2)
\label{eq:Htotal}\end{eqnarray}
The generator of $\kappa_-$ and $\kappa_+$ gauge symmetries 
and reparametrization is:
\begin{eqnarray}
G =&\epsilon_e\pi_e + {\epsilon_p\over 2}(p^2+Z^2) 
+ \{\bar\pi_-\pslash - Z\bar\pi_+ + i \thetabar_+(p^2+Z^2)\}\kappa_- 
\CR 
&+ \{ \bar\pi_-  + i \thetabar_+ \pslash  + i Z \thetabar_{-} 
 -i \bar\rho_- + \bar\zeta_+\pslash  \} \kappa_+ 
\label{eq:Generator+-}\end{eqnarray}
and the phase space action is:
\begin{eqnarray}
S = \int^{\tau_f}_{\tau_i} d\tau ~~\{
 p_\mu{\dot x}^\mu +\pi_e{\dot e} + \bar\pi_+{\dot \theta}_- 
+ \bar\pi_-{\dot \theta}_+ + \bar\rho_-{\dot \zeta}_+ 
\nonumber 
\end{eqnarray}
\vskip -1cm
\begin{eqnarray}
+ { e \over 2}(p^2+Z^2) - \lambda_e\pi_e - \half\lambda_p (p^2+Z^2) 
 - \bar T_+\lambda_- - \bar T'_-\lambda_+ ~~\}  
\label{eq:Action-extended}\end{eqnarray}
The $\kappa_-$ and $\kappa_+$ transformations generated by the
generator G in Eq.(\ref{eq:Generator+-}) are given by:
\begin{eqnarray}
\delta x_\mu = ( \bar\pi_-\Gamma_\mu + 2 i p_\mu\thetabar_+ ) \kappa_-
+( i \thetabar_+ + \bar\zeta_+)\Gamma_\mu\kappa_+ 
\ \ \ , \ \ \ \delta p_\mu = 0 \nonumber
\end{eqnarray}
\begin{eqnarray}
\delta\theta_+ = \pslash\kappa_- + \kappa_+
\ \ \ , \ \ \ \delta\theta_- = -Z\kappa_- \nonumber
\end{eqnarray}
\begin{eqnarray}
\delta\bar\pi_+ = - Z i \bar \kappa_+
\ \ \ , \ \ \  \delta\bar\pi_- = - i (p^2+Z^2)\bar\kappa_- + 
 i \bar\kappa_+\pslash 
\nonumber
\end{eqnarray}
\begin{eqnarray}
\delta\zeta_+ =
 -i \kappa_+ \ \ \ , \ \ \  \delta\bar\rho_- = \bar\kappa_+\pslash 
\label{eq:transf+-}\end{eqnarray}
The action in Eq.(\ref{eq:Action-extended}) is invariant under these transformations if
supplemented also by:
\begin{eqnarray}
\delta\lambda_- =\dot\kappa_- 
\ \ \ , \ \ \  \delta\lambda_+ = \dot\kappa_+ \ \ \ , \ \ \
 \delta\lambda_p = 4 i (\bar\kappa_-\lambda_+ -\bar\kappa_+\lambda_- + 
\bar\kappa_-\pslash\lambda_-)
\nonumber
\end{eqnarray}
as well as invariant under reparametrization
\begin{eqnarray}
\delta x_\mu = p_\mu\epsilon_p
\ \ \ , \ \ \delta p_\mu = 0  \ \ , \ \ 
\delta e = \epsilon_e \ \ , \ \ \delta\pi_e=0 \nonumber
\end{eqnarray}
\begin{eqnarray}
\delta\lambda_e = \dot\epsilon_e \ \ , \ \ 
\delta\lambda_p = \epsilon_e + \dot\epsilon_p 
\ \ , \ \  \epsilon_p(\tau_i)=\epsilon_p(\tau_f)=0    
\label{eq:transf-reparametrization}\end{eqnarray}
In Eq.(\ref{eq:Action-extended}) the bosonic ($\lambda_e , \lambda_p$) Lagrange multipliers
and the Majorana-Weyl ($\lambda_- , \lambda_+$) Lagrange multipliers
are associated with the bosonic and fermionic first
class constraints $\pi_e = p^2 +Z^2 = 0 , \bar T_- =
\bar T_+ = 0 $.

One notices that in the new phase space action of
Eq.(\ref{eq:Action-extended}) only the linear combination 
$-i\bar\rho_- +\bar\zeta_+\pslash$
of new fermionic degrees of freedom appears. The orthogonal combination
does not appear in the action and is thus decoupled from any dynamics of
the system. This "Batalin-Fradkin decoupling" (see refs.
\cite{FADDE}-\cite{FEINB}) 
assures that the correct independent degrees of freedom
defines the extended symmetric system. Namely, we started with
($64-16\times 2-16=$) 16 fermionic degrees of freedom in phase space,
32 degrees of freedom were added and the $\kappa_+$ 
symmetry was introduced.
We have now $(64+32-16\times2-16\times2=16+16)$ 
16 independent degrees of freedom
as in the original system while the other 16 are the "Batalin-Fradkin
decoupled" degrees of freedom.
In the extended symmetry system, in addition to the possible gauge fixing 
(e.g. \cite{KALLO}\cite{KAMIM}) that eliminates the
$\theta_-$ degrees of freedom by fixing the $\kappa_-$ gauge, other
gauge fixings are acceptable as well.
Clearly, as seen in Eq.(\ref{eq:transf+-}), a properly chosen 
gauge fixing ( "unitary" gauge fixing) of the new $\kappa_+$ symmetry
will eliminate the linear combination of the 
new fermionic degrees of freedom 
$-i\bar\rho_- +\bar\zeta_+\pslash$.  
For example a possible unitary
gauge fixing is 
\begin{eqnarray}
\theta_-=0  
{\rm ~~~~~~~~and~~~~~~~~~}  -i\bar\rho_-+\bar\zeta_+\pslash=0
\label{eq:Unitary gauge}\end{eqnarray}
This results in the
same gauge fixed system that was used in \cite{KALLO}. 
A different, interesting, 
gauge fixing that eliminates the old degrees of freedom and leaves
only the new 16 degrees of freedom is simply,
\begin{eqnarray}
\theta_-=0 {\rm ~~~~~~~~and~~~~~~~~~} \theta_+=0   
\label{eq:different gauge}\end{eqnarray}
The gauge fixed D0 system is given in this
gauge in terms of $-i\bar\rho_- +\bar\zeta_+\pslash$ only.  
As in the case of the unitary gauge in Eq.(\ref{eq:Unitary gauge}), the 
Poisson bracket matrix $[\bar T_{\pm \alpha},\chi_{\pm\beta}] $ 
between the constraints $\bar T_{+ \alpha}, \bar T'_{- \alpha}$ 
and the gauge fixing conditions $\chi_-=\theta_-, \chi_+=\theta_+$ 
is not singular since $p^2+Z^2=0$.
Of course, other combinations of $\kappa_-$ and $\kappa_+$ gauge fixings
are also possible.

An interesting set of constraints is defined by:
\begin{eqnarray}
\bar T '_{-\alpha}=&& \bar T_{2\alpha} + 
\bar\Phi_{-\beta}\omega_{\beta\gamma}V_{\gamma\alpha} = 
\bar\pi_-  + i \thetabar_+ \pslash 
 + i Z \thetabar_{-} 
 -i \bar\rho_- + \bar\zeta_+\pslash  \\  \nonumber 
\bar T'_{+\alpha}&\!\!\!=&\!\!\!\bar T_{+} + i 2\left({{p^2+Z^2}\over {p^2}}\right)
\left(\bar\Phi({{1+\Gamma^{11}}\over {2}})\pslash\right)_\alpha 
\\  \nonumber
&\!\!\!=&\!\!\! \bar\pi_-\pslash  - Z\bar\pi_+ + i \thetabar_+(p^2+Z^2)
 +\bar\zeta_+(p^2+Z^2) -i \bar\rho_-
\pslash\left({{p^2+Z^2}\over{p^2}}\right)
\label{eq:Interesting Constr.}\end{eqnarray}
These constraints satisfy the following Poisson bracket relations:
$$
[\bar T'_{+\alpha},\bar T'_{+\beta}] 
=
 -2 i ({{Z^2}\over{p^2}})(p^2+Z^2)\left(\Gamma^0
({ {1+\Gamma^{11}}\over {2}} ) \pslash\right)_{\alpha\beta} \ \ , 
$$
\begin{equation}
[\bar T'_{-\alpha},\bar T'_{-\beta}] 
=
 0 \ \ , 
\ \ \ 
[\bar T'_{+\alpha},\bar T'_{-\beta}] = 0 
\label{eq:Interest. PB}
\end{equation}

We note  in the  $p^2 \gg Z^2$ limit, $\bar T'_{+\alpha}$ and 
$\bar T'_{-\alpha}$ are functions of $( \pi_+ , \theta_- )$ and
$(\pi_- -i \rho_- \ , \ \theta_+ -i \zeta_+)$ only.
It is expected, in this limit, that the system behaves 
as the N=2 massless superparticle  - a system with
16 independent fermionic degrees of freedom in its
phase space, as seen also
directly from the action in Eq.(\ref{eq:Action-d11}).
Indeed, one notes that not only $\bar\rho_-$ and $\zeta_+$
appear only in the linear combinations
$\bar\rho_- + i \bar\zeta_+\pslash$ but now also 
$\bar\pi_- + i \bar\theta_+\pslash$ is the only linear combination
of $\bar\pi_- $ and $\theta_+$ that appears in the constraints.
Thus, after taking into account the decoupling of their
orthogonal linear combination and the fact that
the fermionic degrees of
freedom in phase space are now constrained  by 16 first class
constraints $(\bar T'_{-\alpha})$ while $(\bar T'_{+\alpha})$ are now
second class only (since $p^2+Z^2\neq 0$), one finds indeed in the
$p^2 \gg Z^2$ limit only
16 independent fermionic degrees of freedom as for 
the N=2 massless superparticle. Namely, $64+32-16\times 2 -16 = 
16+16+16$
where the last 16+16 fermionic degrees of freedom are decoupled
in the same sense as the "Batalin Fradkin decoupling" (do not
appear in the constraints or in the Hamiltonian of the extended system).

The  path integral formulation\cite{FRADK} of the system
in Eqs.(\ref{eq:T+-NEW PB}-\ref{eq:Htotal}) with 
$\kappa_-$ and $\kappa_+$ symmetry which has 
only first class constraints is given by:
\begin{eqnarray}
S = \int^{\tau_f}_{\tau_i} d\tau\!\!\!\!\!\!\!&&\{
 p_\mu{\dot x}^\mu + \bar\pi_+{\dot \theta}_- 
+ \bar\pi_-{\dot \theta}_+ + \bar\rho_-{\dot \zeta}_+ + 
\bar\pi_{\lambda+}{\dot \lambda}_- 
+ \bar\pi_{\lambda-}{\dot \lambda}_+
+ \pi_{p}{\dot \lambda}_p         \CR
&&+ \tCC\dot\CP + \tCP\dot\CC + \bar\tCC_+\dot\CP_- +
\bar\tCP_+\dot\CC_- + \bar\tCC_-\dot\CP_+ + \bar\tCP_-\dot\CC_+
 - H_0 \CR
&&+\pi_p\chi + \bar\pi_{\lambda-}\chi_+ + \bar\pi_{\lambda+}\chi_- 
 -{\lambda_p \over 2}(p^2 + Z^2) -\bar T_+\lambda_-  
 -\bar T_-\lambda_+  \CR
&&+ \bar\tCC_+[ \chi_- , \bar T_+ ]\CC_-
+ \bar\tCC_-[ \chi_+ , \bar T_- ]\CC_+ -\tCP\CP  \CR
&&-\tCP_+\CP_- - \tCP_-\CP_+
 - 4\tCP\bar\lambda_-\pslash\CC_- + 4 Z\tCP(\bar\lambda_-\CC_+
 - \bar\lambda_+\CC_-) ~\}
\label{eq:Action-ghosts}\end{eqnarray}

Here, $\CC_\pm$ and $\tCP_\mp$ are canonical pairs of bosonic ghosts
and $\CP_\pm$ and $\tCC_\mp$ are canonical pairs of bosonic anti-ghosts,
associated with the fermionic constraints $T_+$ and $T_-$.
The Majorana-Weyl $\pi_{\lambda+} , \pi_{\lambda-}$  are
the canonical conjugates of the Lagrange multipliers $\lambda_- , 
 \lambda_+$. The bosonic $\pi_p$ is the canonical conjugate of
 the Lagrange multiplier $\lambda_p$ associated with the constraint
$p^2+Z^2=0$ and  $\chi_+$ , $\chi_+$ are gauge fixings. 
The fermionic ghost and its canonical
conjugate are denoted by  $\CC$ and $\tCP$,  and 
the canonical pairs of fermionic 
anti-ghosts as $\CP$ ,$\tCC$ . 

The last 3 lines in Eq.(\ref{eq:Action-ghosts}) are given by:
$
 - [\Psi,\Omega]
$
where the BRST operator $\Omega$ is given by: 
\begin{eqnarray}
\Omega =&& \CP\pi_p + \bar\CP_+\pi_{\lambda-} + \bar\CP_-\pi_{\lambda+}
 + \bar T_+\CC_- + \bar T_-\CC_+ \CR
&&+ {\CC\over 2}(p^2+Z^2)
+2\tCP\bar\CC_-\pslash\CC_- + 2 Z \tCP (\bar\CC_-\CC_+ - 
\bar\CC_+\CC_- ) 
\label{eq:BRST}\end{eqnarray}
and the gauge fixing $\Psi$ is given by:
\begin{eqnarray}
\Psi = -\tCP\lambda - \bar\tCP_+\lambda_-  - \bar\tCP_-\lambda_+
+\tCC\chi +\bar\tCC_+\chi_- +\bar\tCC_-\chi_+ 
\label{eq:Gauge Fixing}\end{eqnarray}

The above $\kappa_-$, $\kappa_+$ symmetric D0 defined in the 
extended phase space $(\theta_\pm , \pi_\mp , \zeta_+ ,\rho_-)$
is physically equivalent to the ordinary D0 with $\kappa_-$ 
symmetry of Eq.(\ref{eq:Action-compactified}). 
This, as mentioned, is demonstrated
by choosing the "unitary" gauge fixing $\chi_\pm$  
in Eq.(\ref{eq:Unitary gauge}) 
that sets the
extended phase space variables $\bar \rho_- + i \bar \zeta_+\pslash$ to zero.
On the other hand the above symmetric system accepts many different
gauge fixings $\chi_\pm$ giving different presentations
of the D0 brane ( for example Eq. (\ref{eq:different gauge}) ).

\eject

{ \bf  D1 brane with $\kappa_-$ and  $\kappa_+$ extended  symmetry}
\vspace{.5cm}

Following along similar lines we present now the extension of this
derivation to the case of a D1 brane. It results in a system with
 $\kappa_-$ and  $\kappa_+$ symmetry which will be discussed below.

The action of the D1 brane consists of the Born-Infeld-Nambu-Goto 
term 
and the Chern-Simons two form $\Omega_2$ term \cite{AGANA} 
\begin{eqnarray}
S = \int {\cal L}(\sigma) d^2\sigma 
= - T\left\{ \int d^2\sigma \sqrt{-{\rm det}( G_{\mu\nu}+{\cal F}_{\mu\nu})} 
+\int \Omega_2 \right\}
\label{eq:D1action}\end{eqnarray}
where $G_{\mu\nu}$ is the supersymmetric induced world-volume metric
and ${\cal F}_{\mu\nu}$ is the supersymmetric Born-Infeld field strength:
\begin{eqnarray}
G_{\mu\nu}&\!\!\!=&\!\!\!\Pi^m_\mu\Pi_{\nu m}
\ \ \ , \ \ \
\Pi^m_\mu=\del_\mu x^m - \bar\theta\Gamma^m\del_\mu\theta \ \ , \ \
\mu,\nu = 0,1 \ ; \ m=0,1,2 . . 9 \\
&\!\!\!&\!\!\!{\cal F}_{01}=F_{01}-b_{01}(\tau_3) \ \ \ , \ \ \ 
F_{\mu\nu}=\del_\mu A_\nu - \del_\nu A_\mu \ \ ,
\end{eqnarray}
\begin{equation}
 b_{01}(\tau_k) = -\bar\theta\Gamma_m\tau_k\left(
\del_0\theta\Pi^m_1 - \del_1\theta\Pi_0^m
+\half(~\del_0\theta(\bar\theta\Gamma^m\del_1\theta) -
\del_1\theta(\bar\theta\Gamma^m\del_0\theta)~) \right)
\label{eq:b01}
\end{equation}
where 
$\theta_\alpha^A , \ \ \alpha=1,2, \cdots, 32$ 
are two Majorana-Weyl spinors with the same chirality, and 
 $\tau_k$ are Pauli matrices acting on indices $A=1,2$. 
The Lagrangian can be rewritten as 
\begin{equation}
{\cal L}(\sigma) 
= - T
\left\{ \sqrt{ G_{01}^2 - G_{00}G_{11} -{\cal F}_{01}^2}  + b_{01}(\tau_1) 
\right\}
\label{eq:D1lagrangian}
\end{equation}

The canonical momenta for the world sheet gauge field is given by the 
electric 
field $E^\mu$ 
\begin{eqnarray}
E^0 = {{\del{\cal L}}\over {\del \dot A_0}} = 0 \ \ , \ \
E^1 = {{\del{\cal L}}\over {\del \dot A_1}} =
{{T{\cal F}_{01}}\over{\sqrt{ G_{01}^2 - G_{00}G_{11} -{\cal F}_{01}^2}}}
\label{eq:E1}\end{eqnarray}
The canonical momenta $\bar\pi_\alpha$ and $p_m$ 
are defined for $\theta_\alpha$ and $x^m$ respectively 
$$
p_m = \tilde p_m - \bar\theta\Gamma_m T_E \del_1\theta 
 \ \ \ \ ,\ \ \ \ 
 \tilde p_m = T{  G_{11}\Pi_{0m}
-G_{01}\Pi_{1m} \over \sqrt{G_{01}^2-G_{00}G_{11}-{\cal F}_{01}^2} }
$$
\begin{equation}
\bar\pi = \bar\theta \Pislash_1 T_E 
-\bar\theta\tilde{\pslash}+
(\thetabar\Gamma^m\del_1\theta)(\thetabar\Gamma_m T_E)
 \ \ \ \ ,\ \ \ \ 
T_E = E^1\tau_3 +T\tau_1 
\label{eq:fermionicmomentum}
\end{equation}
We will suppress the indices $A=1, 2$ of $\theta_\alpha^A$ 
when it is easily recognized. From Eq.(\ref{eq:D1lagrangian}) one finds 
the fermionic constraints 
$\bar\Phi_\alpha^A$ 
\begin{equation}
\bar\Phi_\alpha = \bar\pi_\alpha + (\thetabar\pslash)_\alpha
-(\thetabar\Gamma^m T_E)_\alpha(\del_1 x_m)
+ (\thetabar\Gamma^m\del_1\theta)(\thetabar\Gamma_m T_E)_\alpha ~=0
\label{eq:D1constraints}
\end{equation}
which satisfy 
the Poisson bracket relations:
\begin{equation}
[\bar\Phi_\alpha(\sigma), \bar\Phi_\beta(\sigma')]=2\left( 
(\Gamma^0 {
\tilpslash})_{\alpha\beta} - 
(\Gamma^0 \Pislash_1 T_E)_{\alpha\beta} \right) \delta(\sigma-\sigma')
\label{eq:D1constraintsPB}
\end{equation}

In addition to the fermionic constraints in Eq.(\ref{eq:D1constraints}) one 
finds from Eq.(\ref{eq:D1lagrangian}) also the bosonic first class 
constraints :
\begin{eqnarray}
 \tilde p^2 + G_{11}(E_1^2+T^2) = 0 \ \ , \ \ \tilde p_m\Pi^m_1 = 0
\end{eqnarray}

The constraints in Eq.(\ref{eq:D1constraints}) can be separated covariantly 
into first class and second class 
constraints \cite{KALLO}, \cite{KAMIM} 
:
\begin{equation}
\bar T_{1\alpha} = \left( \bar\Phi (\tilpslash - \Pislash T_E)
({{1+\tau_3}\over{2}}) \right)_\alpha 
\ \ , \ \ \bar T_{2\alpha}= \left( \bar\Phi 
({{1-\tau_3}\over{2}}) \right)_\alpha  
\label{eq:T1T2}
\end{equation}

The Poisson bracket  $[T_{1\alpha},T_{1\beta}]$ vanishes on the constraints 
hyperplane.

These 16 first class constraints $T_{1\alpha}$ generate the local 
$\kappa$ symmetry of the D1 brane. On the other hand:
\begin{eqnarray}
[\bar T_{2\alpha},\bar T_{2\beta}]= 2(\Gamma^0\Pslash\tau_-)_{\alpha\beta}
 \delta(\sigma-\sigma') \end{eqnarray}
{where}
\begin{eqnarray}
P_m =\tilde p_m +E^1\Pi_{1m} = p_m + \thetabar\Gamma_m T_E\del_1\theta 
+ E^1(\del_1x_m - \thetabar\Gamma_m\del_1\theta) 
\nonumber
\end{eqnarray}

Since ~~$ P^2 = \tilde p^2 + 2E_1(\tilde p\Pi_1)+E_1^2G_{11} = -T^2G_{11}$ 
on the constraints hyperplane, we obtain a 
nonvanishing ${\rm det}[\bar T_{2\alpha},\bar T_{2\beta}]$  
(apart from the case $G_{11}=0$)
implying that $\bar T_{2\alpha}$ are 16 second class constraints.
The condition  $G_{11} \neq 0$ is essential for separating the first and
second class constraints and the covariant quantization of the D1 system.
In Ref.\cite{KALLO} it has been emphasized that 
in the static gauge (where $x^\mu=\sigma^\mu$ for $\mu=0,1$) indeed
$G_{11} \neq 0$.
The implications of this fact on the
ground state spectrum and on the relation to the work of 
Ref.\cite{AGANA} on the type IIB fundamental string have been cleared there.
Both Refs.\cite{AGANA} and \cite{KALLO} discuss the properties of the static
gauge and elucidate its physics content.  
Since the static gauge is a natural gauge for D1, we follow this point of 
view. 

We define now a new system in an extended phase space that includes
in addition to the 64 fermionic degrees  of freedom 
$\theta_\alpha^A$ and $\pi_\alpha^A$ extra fermionic 32 degrees
of freedom \cite{FADDE}\cite{BATAL} that satisfy 
\begin{eqnarray}
[\bar \rho_{\alpha}^A(\sigma), \zeta_{\beta}^B(\sigma')]
= \delta(\sigma-\sigma')
\tau_-^{AB}\delta_{\alpha\beta}
\end{eqnarray}

The constraints of the new system 
$\bar {T'_\alpha}^A (x,p,\theta,\pi,\zeta,\rho)$ are obtained from the
constraints in Eq.(\ref{eq:T1T2}) in a similar way the constraints 
in the extended 
phase space in Eq.(\ref{eq:New Constraints}) 
were obtained for the Dparticle. 
Namely, 
$\bar T'^A_{1 \alpha} (x,p,\theta,\pi,\zeta,\rho)= 
\bar T_{1\alpha}^A (x,p,\theta,\pi)$ 
is left unchanged and does not depend on $(\zeta,\rho)$ whereas the other 
constraint $T_{2\alpha}^A$ is modified as 
\begin{eqnarray}
\bar T'^A_{2\alpha} (x,p,\theta,\pi,\zeta,\rho)=
\bar T_{2\alpha}^A (x,p,\theta,\pi)-\bar\rho^A_\alpha 
+ (\bar\zeta^B\Pslash)_\alpha\tau_-^{BA}
\end{eqnarray}
which depends on $(\zeta,\rho)$ and satisfies the Poisson bracket 
relation\footnote{
In deriving Eq.(\ref{eq:newTcommutator}),
the following relations for Majorana $\lambda_i$ have been used
 $(\Gamma_0\Gamma^m T \lambda_2)_\alpha 
(\bar\lambda_3\Gamma_m)_\beta +
(\Gamma_0\Gamma^m T \lambda_2)_\beta 
(\bar\lambda_3\Gamma_m)_\alpha = (\Gamma_0\Gamma_m)_{\alpha\beta}
(\bar\lambda_3\Gamma^m T \lambda_2)$ where $T$ is a matrix in the internal 
space of the Majorana spinors ( such as $\tau_- $ and $T_E$ ). Also:
$
(\bar\lambda_2\Gamma_m)_\alpha
(\bar\lambda_3\Gamma^m T )_\beta
 = (\bar\lambda_2\Gamma_m T)_\alpha
(\bar\lambda_3\Gamma^m )_\beta $
} :
\begin{eqnarray}
[\bar T'_{2\alpha},\bar T'_{2\beta}]
= -2E^1\delta(\sigma-\sigma')
\left( ~2(\Gamma^0\Gamma^m)_{\alpha\beta}(\bar\zeta\Gamma_m\tau_-\del_1\theta) -
(\bar\zeta\Gamma^m)_\alpha (\del_1\bar\zeta\Gamma_m\tau_-)_\beta ~\right) 
\nonumber \\
-2E^1{{\del\delta(\sigma-\sigma')}
\over {\del \sigma'} }(\bar\zeta\Gamma^m)_\alpha
(\bar\zeta\Gamma_m\tau_-)_\beta 
\label{eq:newTcommutator}\end{eqnarray}
In the case of  $E^1=0$ the new system has only first class constraints and
local symmetries $\kappa_1$ and $\kappa_2$ generated by $ T_{1\alpha}$ and
by $ T'_{2\alpha}$ respectively. The symmetric system phase space 
is given by the coordinates 
$\theta_\alpha^A(\sigma) , \pi_\alpha^A(\sigma) ,
\rho_{\alpha}^A(\sigma)$ and $\bar \zeta_{\beta}^B(\sigma)$ 
where the number of independent fermionic degrees of freedom has not
been changed.
Namely, we started with $2 \times 32-2\times 16-16=16$ 
independent fermionic degrees of freedom in phase space and in the extended
phase space we have $3 \times 32-2 \times 32=16+16$(BF) degrees of 
freedom where
the $16$(BF) degrees of freedom are "Batalin Fradkin decoupled" \cite{BATAL}
\cite{FEINB} leaving 16 independent fermionic degrees of freedom.

We note from Eq.(\ref{eq:E1}) 
that setting $E^1=0$ means also that ${\cal F}_{01}=0$
which results in the Lagrangian of Eq.(\ref{eq:D1lagrangian}) 
to be very similar to
the Green-Schwarz (GS) string.

The GS string is described by the action \cite{GREEN} 
\begin{eqnarray}
S = \int {\cal L}(\sigma) d^2\sigma 
= - {T\over 2}\int d^2\sigma \sqrt{h}
h^{\alpha\beta}G_{\alpha\beta} 
+\int {\cal L}_2 d^2\sigma 
\end{eqnarray}
\begin{eqnarray}
 {\cal L}_2 &\!\!\!=&\!\!\! -{T}\epsilon^{\alpha\beta}\del_\alpha x^m
(\thetabar^1\Gamma_m\del_\beta\theta^1 - \thetabar^2\Gamma_m
\del_\beta\theta^2)-{T}\epsilon^{\alpha\beta}
(\thetabar^1\Gamma^m\del_\alpha\theta^1)
(\thetabar^2\Gamma_m\del_\beta\theta^2) 
\CR
&\!\!\!=&\!\!\!-{T}(\del_0x^m
(\thetabar\Gamma_m\tau_3\del_1\theta) - 
\del_1x^m
(\thetabar\Gamma_m\tau_3\del_0\theta) )
-{T\over 2}(\thetabar\Gamma^m\tau_3\del_0\theta)
(\thetabar^A\Gamma_m\del_1\theta^A)  \CR
&\!\!\!&\!\!\! \ \ \ \ \ \ \ \ \ \ \ \ \ \ \ \ \ \ \ \ \ \ \
  \ \ \ \ \ +{T\over 2}(\thetabar\Gamma^m\tau_3\del_1\theta)
(\thetabar^A\Gamma_m\del_0\theta^A) 
\end{eqnarray}
This can be compared to the $b_{01}(\tau_k)$ of Eq.(\ref{eq:b01}) which 
can be
written also as:
\begin{eqnarray}
 b_{01}(\tau_k)&\!\!\!=&\!\!\!(\del_0x^m
(\thetabar\Gamma_m\tau_k\del_1\theta) - 
\del_1x^m
(\thetabar\Gamma_m\tau_k\del_0\theta) )
 +{1\over 2}(\thetabar\Gamma^m\tau_k\del_0\theta)
(\thetabar^A\Gamma_m\del_1\theta^A)  \CR
&\!\!\!&\!\!\! \ \ \ \ \ \ \ \ \ \ \ \ \ \ \ \ \ \ \ \ \ \ \
  \ \ \ \ \ -{1\over 2}(\thetabar\Gamma^m\tau_k\del_1\theta)
(\thetabar^A\Gamma_m\del_0\theta^A) 
\end{eqnarray}
Thus ${\cal L}_2 = -T b_{01}(\tau_3)$ compared to $-T b_{01}(\tau_1)$ in the
Wess-Zumino term of the D1 brane. Similarly, using the equation of motion
for $h^{\alpha\beta}$ one notices that the D1 action in Eq.(\ref{eq:D1action})
with $E_1=0$ (namely ${\cal F}_{\mu\nu}=0$) is identical to the
Green and Schwarz action when $\tau_3$ is replaced by $\tau_1$ \cite{KAMIM}. 
Since we are using the static gauge as a natural gauge for D1 \cite{AGANA}, 
the massless modes are projected out. 
This relation between the physics of the type IIB fundamental string 
and the D1 system in the static gauge has been noted in \cite{KALLO}.

We also note that the electric field $E^1$ is quantized and 
represents the number of fundamental string bound to the D1 brane 
producing $(n, m)$ string 
\cite{WITTEN}, \cite{POLTASI}. 
Therefore we have succeeded to extend the system where all the 
second class constraints are turned into 
first class constraints at least for the case of the $(0, 1)$ string, namely 
the genuine D1 brane without F1 
provided the massless modes are projected out 
by using, for instance, the static gauge.

\vspace{1cm}
{\bf Acknowledgement}
\vspace{.5cm}

This work is supported in part by Grant-in-Aid 
for Scientific Research from the Japan Ministry 
of Education, Science and Culture for 
the Priority Area 291, by the 
Japan Society for Promotion of
Sciences and by the
Israeli Science Foundation. 
The authors thank Kenji Nagami for pointing out the hermiticity properties 
of Grassman variables and Joshua Feinberg for useful discussions.
MM acknowledges the warm hospitality at TIT and
Fermilab where this work has been completed.

\end{document}